\documentstyle[aps,epsfig,prb,multicol]{revtex}

\relax 
\begin{document}
%\preprint{} \draft

\title{Electrodynamics of quasi-two-dimensional BEDT-TTF charge transfer salts}

\author{Stephen Hill\cite{email1}}
\address{Department of Physics, Montana State University, Bozeman, MT 59717}

\date{\today}
%\maketitle
\maketitle

\begin{abstract}

We consider the millimeter-wave electrodynamics specific to
quasi-two-dimensional conductors and superconductors based on the
organic donor molecule BEDT-TTF. Using realistic physical
parameters, we examine the current polarizations that result for
different oscillating (GHz) electric and magnetic field
polarizations. We show that, in general, it {\em is} possible to
discriminate between effects (dissipation and dispersion) due to
in-plane and interlayer ac currents. However, we also show that it
is {\em not} possible to selectively probe any single component of
the in-plane conductivity tensor, and that excitation of
interlayer currents is strongly influenced by the sample geometry
and the electromagnetic field polarization.

\end{abstract}

%\pacs{PACS numbers: 71.18.+y, 71.27.+a, 74.25.Nf}
%\bigskip
%\clearpage
\begin{multicols}{2}[]

This brief report has been motivated by a number of recent studies
of the millimeter-wave electrodynamics of quasi-two-dimensional
(Q2D) organic conductors and superconductors based on the organic
donor molecule BEDT-TTF (bis-ethylenedithio-tetrathiafulvalene, or
ET for short).\cite{dressel,schrama,ardavan,hill1,shibauchi,mola1}
In recent years, measurements of the microwave surface impedance
of high temperature superconductors (HTS) have played an important
role in determining whether the superconducting gap function
possesses nodes at certain points, or along certain lines in
reciprocal space.\cite{nodes} Indeed, microwave penetration depth
measurements are often cited as one of the key pieces of evidence
supporting a $d-$wave scenario in the hole-doped HTS.\cite{dwave}
Similar studies have also been published for several organic
superconductors,\cite{dressel,carrington} though the symmetry of
the superconducting state in these lower T$_c$ compounds remains
the subject of considerable debate.\cite{ross} Aside from studies
of superconductivity, millimeter-wave spectroscopies have also
played a vital role in furthering our understanding of the often
unusual normal state properties of many organic
conductors.\cite{ardavan,hill1,hill2}

Experiments usually involve placing a tiny single crystal at
various locations within either rectangular or cylindrical
enclosed cavities;\cite{klein,mola2} the linear dimensions of a
typical sample are $\sim 0.5-2$mm parallel to the conducting
planes and $\sim 0.05-0.5$mm in the perpendicular direction. In
principle, dissipation due to the sample may be determined by
measuring the change in dissipation (i.e. change in the resonance
$Q-$factor) within the cavity upon insertion of the sample, while
the penetration depth is related to the change in the central
resonance frequency ($f_o$) of the cavity upon insertion of the
sample. In practice, this rigorous approach is rarely followed;
rather, changes in dissipation and dispersion are recorded whilst
varying some other external parameter, e.g. magnetic field,
temperature, etc. The latter approach is particularly well suited
to magnetic resonance experiments, e.g. cyclotron
resonance,\cite{ardavan,hill1} Josephson plasma
resonance,\cite{mola1,shibauchi} etc., where one is not so
interested in the absolute value of the power dissipation within
the cavity, but rather in the magnetic field strength at which
maximum dissipation occurs. Furthermore, one can perform fairly
reliable lineshape analysis in this way, provided suitable care is
taken to truly separate dissipative and dispersive contributions
to the measurement.\cite{mola2}

The high degree of anisotropy in the Q2D ET salts results in a
situation in which the in-plane and interlayer electrodynamics
differ considerably, particularly at low temperatures; by in-plane
and interlayer electrodynamics, we mean the dynamics of in-plane
and interlayer ac currents. These differences are attributable to
the 3 to 4 orders of magnitude difference in conductivity parallel
($\sigma_\parallel$) and perpendicular
($\sigma_\perp\ll\sigma_\parallel$) to the conducting layers, as
discussed at length in ref. [11].\nocite{hill2} In this paper, we
consider the experimental geometries employed by several groups
and, in each case, we discuss contributions to the electrodynamics
due to both in-plane and interlayer currents. Before proceeding,
we wish to correct a common misconception concerning the response
of an anisotropic (super-) conductor to different electromagnetic
field polarizations; namely that it is the polarizations of the
induced currents that govern the electrodynamics, not the
polarizations of the electromagnetic fields. We specifically
examine the current polarizations arising in Q2D conducting ET
salts for various ac electric or magnetic field polarizations.

Case I $-$ in-plane measurements. The most obvious situation
involves placing a sample within an oscillating electric field
({\bf\~ E}) with polarization parallel to its highly conducting
layers. Under these conditions, it is actually the displacement
current $\partial${\bf\~ D}/$\partial t$ in the cavity that is
responsible for driving currents within the sample (this is
discussed further below). At liquid helium temperatures, the
skin/penetration depths ($\delta_\parallel,\lambda_\parallel$) for
in-plane currents in the normal/superconducting states range from
about 0.2 to $5\mu$m in high quality Q2D conducting ET salts.
Thus, for this geometry, in-plane currents flow within a thin
layer at the surface of the sample a distance $\delta_\parallel$
or $\lambda_\parallel$ from the sample's edges and faces, as
illustrated in Fig. 1a. However, the shape of a typical Q2D ET
salt is incompatible with the axial symmetry associated with such
an electric field, i.e. single crystal samples generally form as
thin platelets with the low conductivity direction perpendicular
to the plane of the sample (see Fig. 1). Furthermore, such an
electric field is also incompatible with the axial symmetry
associated with the Q2D conductivity tensor, i.e.
$\sigma_\parallel$ is approximately isotropic within the planes,
while $\sigma_\perp$ is considerably lower perpendicular to the
planes. Consequently, an ac electric field parallel to the layers
will become disrupted in the vicinity of the sample and may
distort the flow of currents within the sample to such an extent
that interlayer currents are also excited (not shown in Fig. 1a).

A better solution for in-plane measurements involves placing the
sample within an oscillating magnetic field ({\bf\~ H}) with
polarization perpendicular to its highly conducting layers.
Although this produces circulating currents, they are confined to
within the highly conducting layers (see Fig. 1b) and, thus, one
only has to consider in-plane electrodynamics. Once again, these
currents only flow within a distance $\delta_\parallel$ or
$\lambda_\parallel$ of the sample's edges and faces. Fringing
fields associated with the demagnetization of the interior of the
sample produce negligible interlayer currents; this has been
verified experimentally\cite{mola2} and is discussed further
below. Although this second method is preferable (Fig. 1b), both
configurations (Figs. 1a and b) reliably probe the in-plane
electrodynamics of Q2D (super-) conducting ET salts. However, the
conductivity $\sigma_\parallel$ (or
$\delta_\parallel$,$\lambda_\parallel$) measured by either method
represents an average over all in-plane directions, i.e. it is
impossible to measure a single component of the in-plane
conductivity tensor.

Case II $-$ interlayer measurements. This situation is far more
complex than the in-plane case. In particular, the electrodynamics
differ considerably depending on whether the interlayer
skin/penetration depth ($\delta_\perp$,$\lambda_\perp$) is smaller
(skin-depth regime) or larger (depolarization regime) than the
linear dimension ($L$) of the sample parallel to the conducting
layers. In the latter case, the simplest experimental
configuration involves placing the sample within an oscillating
electric field ({\bf\~ E}) with its polarization perpendicular to
the highly conducting layers (see Fig. 1c). Under such
circumstances, the  {\bf\~ E}-field penetrates the sample
uniformly and drives bulk interlayer ac currents between the flat
platelet surfaces; dissipation is then due entirely to
$\sigma_\perp$.

The above situation is rarely, or never, realized at liquid helium
temperatures in typical conducting ET salts $-$ the reason is
because $\delta_\perp$ and $\lambda_\perp$ rarely exceed about
$200\mu$m (usually $<100\mu$m). Consequently, even for the
smallest samples studied, any induced interlayer currents will
attenuate appreciably from the edges towards the center of the
sample. If we assume a quasi-static approximation, which is
justified on the grounds that the radiation wavelength (several
millimeters) is larger than a typical sample, it becomes apparent
that there is no ac electric or magnetic field polarization that
can excite pure interlayer currents. To illustrate what happens in
the skin-depth regime we first consider the case with the ac
electric field ({\bf\~ E}) polarization perpendicular to the
conducting layers. Assuming $\delta_\perp\approx L/10$, which is
typical of many of the published studies, screening of the
electric field within the bulk of the sample requires a uniform
surface charge density to alternately flow between the large flat
faces of the sample, as illustrated in Fig. 2. The only way that
this charge can complete the circuit from one face to the other is
via the edges of the sample, i.e. within a distance $\delta_\perp$
or $\lambda_\perp$ of the perimeter of the platelet. Consequently,
currents must flow across the flat sample faces (i.e. parallel to
the high conductivity planes) in order to maintain the uniform
surface charge density necessary to screen the electric field from
within the interior of the sample. These in-plane currents will be
confined to a depth $\delta_\parallel$ or $\lambda_\parallel$
beneath the flat sample surfaces (Fig. 2). One can think of this
current flow as a continuation of the displacement currents
surrounding the sample; these, in turn, complete a circuit with
the surface currents in the walls of the cavity.

One other way to generate interlayer currents is to place the
sample within an oscillating magnetic field ({\bf\~ H}) with its
polarization parallel to the highly conducting layers (see Fig.
3). In spite of the conflicting symmetries associated with this
field and the Q2D electronic system, this configuration seems to
be the most widely used for studies of Q2D conducting ET
salts.\cite{hpar} As in the previous case (Fig. 2), a combination
of interlayer and in-plane currents are necessary to screen the
oscillatory {\bf\~ H}-field from within the sample. This time, the
currents complete a circuit within the sample (i.e. there is no
displacement current) that curl in a left-handed sense around the
oscillatory {\bf\~ H}-field. In-plane currents flow in opposite
directions across the flat sample faces within a surface layer of
thickness $\delta_\parallel$ or $\lambda_\parallel$. Meanwhile,
interlayer currents complete the circuit at the sample edges
within a surface layer of thickness $\delta_\perp$ or
$\lambda_\perp$. This situation is depicted in Fig. 3b.

Next we consider the relative contributions to the dissipation
within the sample due to interlayer and in-plane currents for case
II above (Figs. 2 and 3). In the skin-depth regime, dissipation is
governed by surface resistance {\bf R}$_S$, i.e. the real part of
the surface impedance

\begin{equation}
{\bf \hat Z}_S  = {\bf R}_S  + i{\bf X}_S  = \sqrt {\frac{{i\mu
_{\rm o} {\rm \omega }}}{{\sigma _1  - i\sigma _2 }}}
 \end{equation}

\noindent{Provided one is not in the anomalous skin effect regime
($\delta$ $\ll$ mean free path), which is never the case in even
the best quality samples,\cite{anomalous} the above expression may
be expanded assuming a Drude form for the
conductivity.\cite{klein} In doing so, one finds two solutions for
{\bf R}$_S$,}

\begin{equation}
{\bf R}_S  = \frac{{\mu _o {\rm \omega \delta }_{\rm o} }}{{\rm
2}}\left( {1 - \frac{{{\rm \omega \tau }}}{{\rm 2}} +
\frac{{\left( {{\rm \omega \tau }} \right)^2 }}{{\rm 8}} + ....}
\right)
\end{equation}

\noindent{and}

\begin{equation}
{\bf R}_S  = \frac{{\mu _o {\rm \omega \delta }_{\rm o} }}{{\rm
2}}\left( {\frac{1}{{\sqrt {{\rm \omega \tau }} }} - \frac{{\rm
1}}{{2\left( {{\rm 2\omega \tau }} \right)^{{\raise0.7ex\hbox{$5$}
\!\mathord{\left/
 {\vphantom {5 2}}\right.\kern-\nulldelimiterspace}
\!\lower0.7ex\hbox{$2$}}} }} + ....} \right),
\end{equation}

\noindent{depending on whether i) $\omega\tau<1$ [Eq. (2),
Hagen-Rubens regime, $\sigma_1>\sigma_2$], or ii) $\omega\tau>1$
[Eq. (3), Relaxation regime, $\sigma_2>\sigma_1$]; $\delta_o$ is
the penetration depth obtained from the dc conductivity, i.e.
$\delta_o=(\sigma_{dc}\omega\mu_o/2)^{-0.5}$. Upon application of
a dc magnetic field, the $\omega\tau$ product may be replaced by
$(\omega-\omega_c)\tau$, where $\omega_c$ is the cyclotron
frequency $(= eB/m^*)$. In general, the most relevant case to
compare with published low temperature data is the latter
(relaxation regime), though this situation is complicated when a
dc magnetic field is included.\cite{dcfield} In either case,
however, {\bf R}$_S$ is proportional to $\delta_o$. Thus, in the
normal metallic state, dissipation due to the respective current
polarizations will be proportional to the appropriate skin depth
($\delta_\parallel$ or $\delta_\perp$) multiplied by an
appropriate area ($a_\parallel$ or $a_\perp$) for the surface
across which that current flows. Then the ratio of the power
dissipation due to interlayer ($P_\perp$) and in-plane
($P_\parallel$) currents is given by
$P_\perp/P_\parallel=a_\perp\delta_\perp/a_\parallel\delta_\parallel$.}

The skin-depth ratio $\delta_\perp/\delta_\parallel$ is a
parameter which is often used as a measure of anisotropy, and in
the case of the Q2D conducting ET salts ranges from about 30 to
over 100. Thus, we see that interlayer currents will tend to
dominate the measured dissipation, unless an extremely thin sample
is used, i.e. area of faces exceeds the area of the edges by a
compensating factor of 30 to 100. In reality, the aspect ratio
$a_\parallel/a_\perp$ is typically in the range 1 $-$ 5 for the
samples used in most published studies (see e.g. [2,4$-$6]).
Indeed, this is often cited as the reason why one can confidently
attribute dissipation as being due entirely to $\sigma_\perp$ for
measurements with {\bf\~ H} parallel to the conducting layers
(again, see e.g. [2$-$6]). This has been confirmed by the few
studies in which both the in-plane (case I) and interlayer (case
II) electrodynamics have been probed independently, e.g. in ref.
[13], different cyclotron resonance lineshapes are observed for
{\bf\~ H} parallel and perpendicular to the conducting layers
which may be explained in terms of the different dominant current
polarizations within the sample.

Things become more extreme in the superconducting phase, where the
anisotropy parameter $\gamma$ ($=\lambda_\perp/\lambda_\parallel$)
is believed to exceed 100 (and maybe up to 200) in the most widely
studied $\kappa-$phase ET salts. In this case, it is expected that
$\sigma_2\gg\sigma_1$, and that {\bf R}$_S$ is given approximately
by

\begin{equation}
{\bf R}_S  = \frac{{\sigma _1 }}{{2\sigma _{\rm 2} }}\sqrt
{\frac{{\mu _o {\rm \omega }}}{{\sigma _{\rm 2} }}}.
\end{equation}

\noindent{Matters become even more complicated in the presence of
an external dc magnetic field, due to the creation of a mixed
state.\cite{shibauchi,mola1} The large $\gamma$ parameter results
in a situation in which interlayer currents penetrate well into
this mixed state. Consequently, one can expect the vortex
structure/dynamics to have a major influence on $\sigma_\perp$.
Indeed, the effects of vortices show up in the interlayer
electrodynamics of the 10 K superconductor
$\kappa-$(ET)$_2$Cu(NCS)$_2$, when probed with {\bf\~ H} parallel
to the conducting layers (Fig. 3);\cite{shibauchi,mola1} a plasma
mode is observed $-$ attributable to interlayer Josephson currents
$-$ which is not seen at all for {\bf\~ H} perpendicular to the
layers. This confirms that, even within the superconducting phase,
$\sigma_\perp$ dominates the dissipation in the second case (II)
considered above, i.e. with {\bf\~ H} ({\bf\~ E}) parallel
(perpendicular) to the layers.}

Finally, we note that extreme caution should be exercised when
comparing data for different electromagnetic field polarizations.
In particular, although it is clear that $\sigma_\perp$ dominates
the second case (II) considered above, it is apparent that
different sample aspect ratios $a_\parallel/a_\perp$ are
appropriate depending on whether {\bf\~ H} (Fig. 2) or {\bf\~
E}-field (Fig. 3) excitation is used, and whether the sample is
superconducting or not. Furthermore, for {\bf\~ H}-field
excitation, different aspect ratios will invariably apply for
different polarizations within the conducting planes; this is the
most probable explanation for the azimuthal polarization
dependence observed in ref. [2].

In summary, using realistic sample parameters, we have examined
the electrodynamics appropriate to Q2D conducting and
superconducting ET salts, and for various different
electromagnetic field polarizations. We find that it {\em is}
possible to discriminate between the effects (dissipation and
dispersion) of in-plane and interlayer ac currents. We also show
that the ratios of the areas of different sample faces has a major
bearing on the relative contributions of $\sigma_\perp$ and
$\sigma_\parallel$ to the dissipation.

We are indebted to Neil Harrison and Jochen Wosnitza for
stimulating discussion. This work was supported by the Office of
Naval Research (N00014-98-1-0538).

% References

\end{multicols}

\bigskip

\clearpage
\begin{twocolumn}

\begin{figure}
\centerline{\epsfig{figure=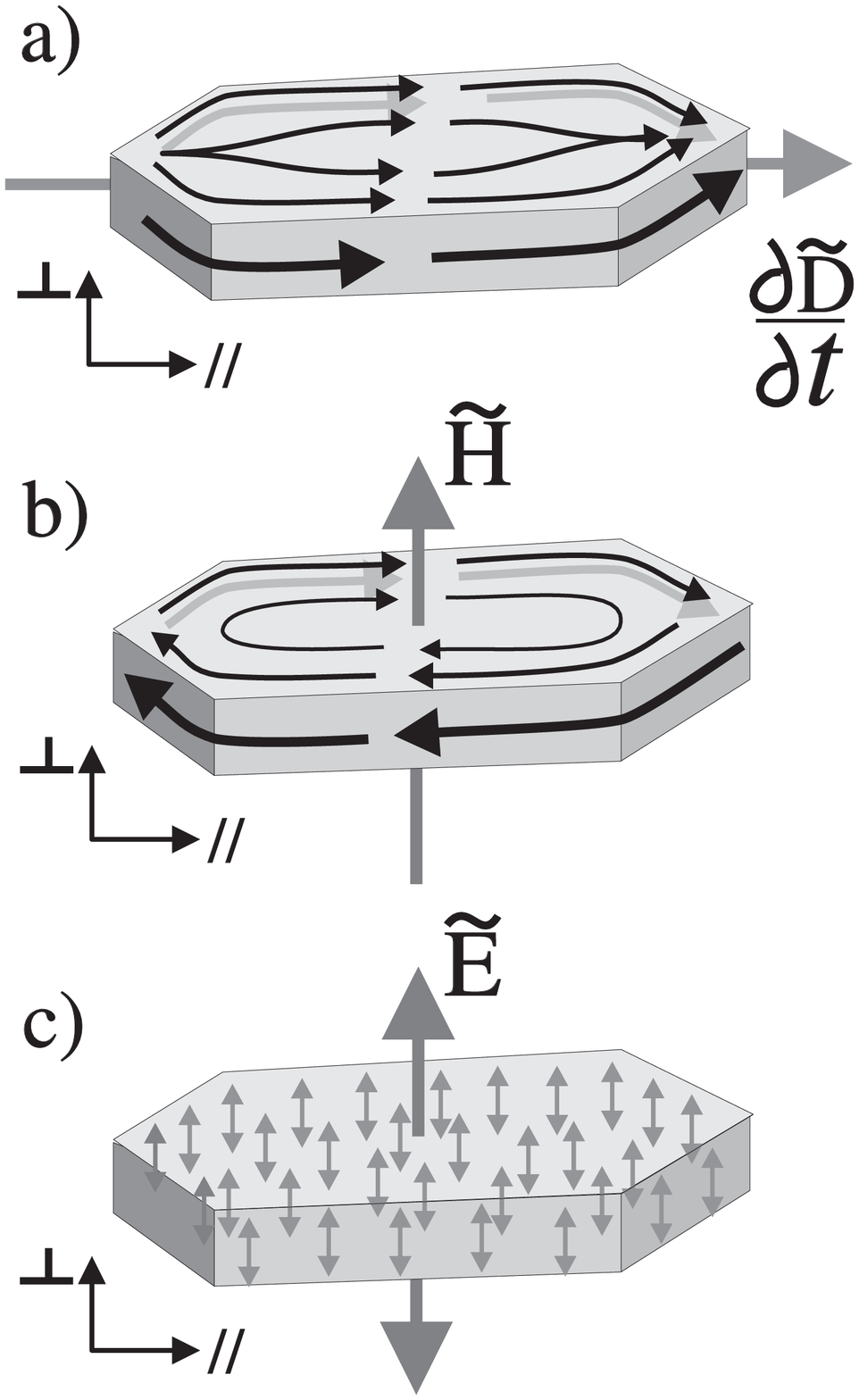,width=55mm}}\bigskip
\caption{Schematic representing the flow of in-plane currents
(case I) for a) {\bf\~ E}-field excitation, and b) {\bf\~ H}-field
excitation; in both a) and b), currents flow only within a
distance $\delta_\parallel$ or $\lambda_\parallel$ $(\ll$ sample
dimensions) of the sample surface. c) depicts uniform interlayer
currents (case II) for a sample in the depolarization regime,
driven by an {\bf\~ E}-field perpendicular to the layers.}
\label{Fig. 1}
\end{figure}

\bigskip
\bigskip
\bigskip
\bigskip
\bigskip
\bigskip
\bigskip
\bigskip
\bigskip
\bigskip
\bigskip
\bigskip
\bigskip
\bigskip
%\clearpage

\begin{figure}
\centerline{\epsfig{figure=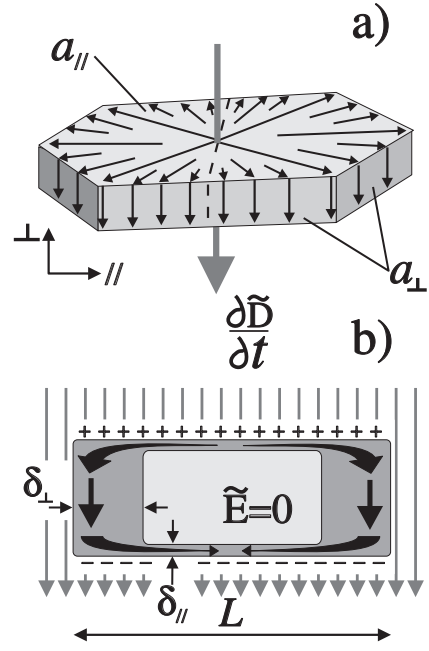,width=55mm}}
\bigskip
\caption{{\bf\~ E}-field excitation of interlayer currents (case
II) for a sample in the skin-depth regime. The currents respond to
the surrounding displacement current $\partial${\bf\~ D}/$\partial
t$ in such a way as to maintain the necessary surface charge to
screen the {\bf\~ E}-field from within the sample. The dashed line
in a) indicates the location of the cross-section shown in b).
Note: both in-plane and interlayer currents are excited; the
former flow within a surface layer of thickness $\delta_\parallel$
parallel to the large sample faces, while the latter flow within a
surface layer of thickness $\delta_\perp$ at the sample edges.}
\label{Fig. 2}
\end{figure}

\bigskip

%\clearpage

\begin{figure}
\centerline{\epsfig{figure=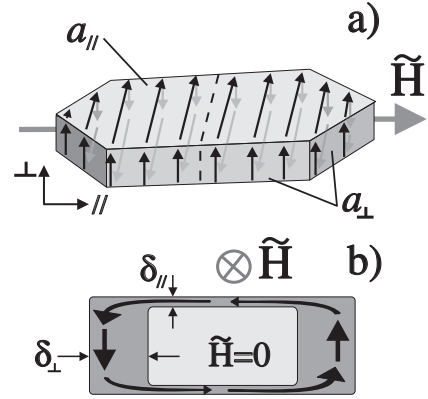,width=55mm}}
\bigskip \caption{{\bf\~ H}-field excitation of interlayer currents (case
II) for a sample in the skin-depth regime. The dashed line in a)
indicates the location of the cross-section shown in b). Note:
both in-plane and interlayer currents are excited; the former flow
within a surface layer of thickness $\delta_\parallel$ parallel to
the large sample faces, while the latter flow within a surface
layer of thickness $\delta_\perp$ at the sample edges.}
\label{Fig. 3}
\end{figure}
\end{twocolumn}
%\end{multicols}


\begin{references}
\bibitem{anomalous}
It is well documented that interlayer transport in the Q2D
conducting ET salts is on the verge of being incoherent; this is
due to the extremely weak band dispersion in the interlayer
direction. Thus, a carrier contributing to in-plane current flow
(within a layer of thickness $\delta_\parallel$) will undergo many
collisions before it leaves the skin layer, i.e. a carrier will
typically only make it across a few conducting layers
($\ll\delta_\parallel$) between collisions. Meanwhile,
$\delta_\perp$ is considerably greater than the mean free path in
even the cleanest ET salts (mean free path $\sim 5\mu$m).
\bibitem{dcfield}
$\sigma_1$ and $\sigma_2$ vary dramatically in the vicinity of the
resonance condition $\omega_c=\omega$, with $\sigma_2$ changing
sign at $\omega_c=\omega$. This subject will be treated in detail
in a future publication.
\end{references}
\end{document}